\documentclass[12pt]{article}

\def\bSig\mathbf{\Sigma}

\usepackage{natbib}
\usepackage{hyperref}
\usepackage{amsmath}
\usepackage{amsfonts}
\usepackage{amssymb}
\usepackage{yhmath}
\usepackage{boldline}
\usepackage{graphicx}
\usepackage{filecontents}
\usepackage{setspace}
\usepackage{booktabs}
\usepackage{todonotes}
\usepackage{changes}
\usepackage{xurl} 
\usepackage{algorithm}
\usepackage[noend]{algpseudocode}
\usepackage{subcaption}
\usepackage{footnote}
\usepackage{threeparttable}

\DeclareMathOperator*{\argmax}{argmax} 

\newcommand{\blind}{0}

\begin{document}

\def\spacingset#1{\renewcommand{\baselinestretch}%
{#1}\small\normalsize} \spacingset{1}


\if0\blind
{
  \title{\bf Finite-Sample Two-Group Composite Hypothesis Testing via Machine Learning}
  \author{Tianyu Zhan\thanks{Tianyu Zhan is an employee of AbbVie Inc. Corresponding author email address: tianyu.zhan.stats@gmail.com.}\hspace{.2cm}\\
    \footnotesize Data and Statistical Sciences, AbbVie Inc. 1 Waukegan Road, North Chicago, IL 60064\\
    and \\
    Jian Kang\thanks{Jian Kang is Professor in the Department of Biostatistics at the University of Michigan, Ann Arbor. Kang’s research was partially supported by NIH grants R01DA048993,  R01MH105561 and R01GM124061 and NSF grant IIS2123777. Email address: jiankang@umich.edu.} \\
    \footnotesize Department of Biostatistics, University of Michigan, 1415 Washington Heights, Ann Arbor, MI 48109}
\date{}
  \maketitle
} \fi

\if1\blind
{
  \bigskip
  \bigskip
  \bigskip
  \begin{center}
    {\LARGE\bf Title}
\end{center}
  \medskip
} \fi

\bigskip
\begin{abstract}
In the problem of composite hypothesis testing, identifying the potential uniformly most powerful (UMP) unbiased test is of great interest. Beyond typical hypothesis settings with exponential family, it is usually challenging to prove the existence and further construct such UMP unbiased tests with finite sample size. For example in the COVID-19 pandemic with limited previous assumptions on the treatment for investigation and the standard of care, adaptive clinical trials are appealing due to ethical considerations, and the ability to accommodate uncertainty while conducting the trial. Although several methods have been proposed to control type I error rates, how to find a more powerful hypothesis testing strategy is still an open question. Motivated by this problem, we propose an automatic framework of constructing test statistics and corresponding critical values via machine learning methods to enhance power in a finite sample. In this article, we particularly illustrate the performance using Deep Neural Networks (DNN) and discuss its advantages. Simulations and two case studies of adaptive designs demonstrate that our method is automatic, general and pre-specified to construct statistics with satisfactory power in finite-sample. Supplemental materials are available online including R code and an R shiny app. 
\end{abstract}

\noindent%
\newline
\newline
\vfill

\newpage
\spacingset{1.5} 

\section{Introduction}
\label{sec:motivation_two_sample}

In simple hypothesis testing, the Uniformly Most Powerful (UMP) level test can be constructed based on Neyman-Pearson Fundamental Lemma \citep{np1933}. For a large class of composite hypothesis testing problems for which a UMP test does not exist, there sometimes exists a UMP test among all unbiased tests, under which no alternative
has the probability of rejection be less than the size of the test \citep{leh2006}. However, even in exponential family, it can be challenging to prove the existence and further construct such UMP unbiased tests in finite-sample. For example in the $\textit{Behrens-Fisher}$ problem of testing equality of the means from two normal distributions with unknown variances and unknown variance ratio, the $\textit{Welch approximate t-solution}$ is one of approximate solutions for practical purposes \citep{wel1951}.

Another important application of composite hypothesis testing is to understand the effect of a treatment relative to the standard of care in randomized clinical trials (RCTs) \citep{chen2014, fda2019, ema2007}. In the COVID-19 pandemic with limited knowledge of the treatment profiles, adaptive designs, for example the Adaptive COVID-19 Treatment Trial (ACTT) \citep{nihcovid}, can be more efficient than traditional RCTs by allowing for prospectively planned modifications to design aspects based on accumulated unblinded data \citep{bre2009, chenet2010}. Despite several proposed statistical methods to control the type I error rate \citep{bau1994, cui1999, bre2009}, how to find a more powerful hypothesis testing strategy remains open. The improved hypothesis testing strategy is especially attractive to patients, because a safe and efficacious drug can be delivered more efficiently and more ethically to meet unfulfilled medical needs.

Motivated by this problem, we propose a general two-stage method via machine learning approaches to conduct two-group composite hypothesis testing when the theoretically most powerful test does not exist or is hard to characterize. We first construct test statistics and then estimate critical values by simulated null samples. Our computational method leverages machine intelligence to establish an automatic framework of performing hypothesis testing with only some basic knowledge of a problem at hand. The proposed method is also general in the sense that it can incorporate existing statistics to seek power improvement if possible. Additionally, our method constructs pre-specified decision function for hypothesis testing before observing current data. This is especially appealing to ensure integrity of adaptive clinical trials as considered in Section \ref{s:real}. 

In this article, we particularly apply Deep Neural Networks (DNN) to the proposed framework due to its strong functional representation and scalability to large datasets \citep{goo2016,cho2018}. Recently, DNN has been applied to hypothesis testing based on the maximum mean discrepancy (MMD) and its variants \citep{che2019, kir2020, kub2020, liu2020} and on testing non-linear effects \citep{liu2017} in large samples, but our aim is to enhance power in a finite sample. More motivations for using DNN are illustrated in Section \ref{sec:second_DNN}, and its advantages are demonstrated by simulations in Web Appendix D. 

The remainder of this article is organized as follows. In Section \ref{sec:method_intro}, we describe the problem setup and the motivation of our method. Then we introduce the proposed two-stage DNN-guided hypothesis testing framework in Section \ref{sec:method}. Simulations are performed to demonstrate advantages of the proposed method in Section \ref{s:sim_eg_1} and \ref{s:sim_eg_2}. We further apply our method to the ACTT on COVID-19 in Section \ref{s:ACTT}, and another adaptive trial MUSEC (Multiple Sclerosis and Extract of Cannabis) in Section \ref{s:MUSEC}. Concluding remarks are provided in Section \ref{s:discussions}. 

\section{Problem setup}
\label{sec:method_intro}

\subsection{Composite hypothesis testing with two groups of data}

Based on the motivating example of the ACTT on COVID-19, we consider a composite hypothesis testing problem using two groups of independent data $\boldsymbol{x}_j$, $j=1, 2$, of equal size $n$ where samples in group $j$ are independent and identically distributed (i.i.d.) with a probability mass function (pmf) or a probability density function (pdf) denoted as  $f(x; \theta_j, \boldsymbol{\eta}_j)$. The parameter of interest $\theta_j \in \mathbb{R}$ is considered as a scaler quantity, while the nuisance parameters $\boldsymbol{\eta}_j \in \mathbb{R}^{w}$ is a vector of dimension $w$ in group $j$. For example in the ACTT, group $1$ is the standard of care, group $2$ is the remdesivir under evaluation, and $\theta_j$ can be the probability of achieving hospital discharge at Day 14 \citep{nihsimple, gil2020} in group $j$, $j=1, 2$. Suppose that the following composite null hypothesis $H_0$ is to be tested agains t a composite alternative hypothesis $H_1$ with a one-sided type I error rate controlled at $\alpha$,
\begin{align}
H_0: & \:\: \theta_1 = \theta_2, \:\:\:\:\:\: \text{versus} \:\:\:\:\:\: H_1: \:\: \theta_1 < \theta_2, \label{H_0_eg}
\end{align}
where a hypothesis is said to be composite if it contains a class of distributions \citep{leh2006}. Under $H_0$ in (\ref{H_0_eg}), the class of distribution is denoted as
\newline $\mathcal{F}_0 = \left\{f(x; \theta_{12}, \boldsymbol{\eta}_1), f(x; \theta_{12}, \boldsymbol{\eta}_2); \theta_{12} \in \mathbb{R}, \boldsymbol{\eta}_1 \in \mathbb{R}^{w}, \boldsymbol{\eta}_2 \in \mathbb{R}^{w} \right\}$; while \newline $\mathcal{F}_1 = \left\{f(x; \theta_{1}, \boldsymbol{\eta}_1), f(x; \theta_{2}, \boldsymbol{\eta}_2); \theta_{1} \in \mathbb{R}, \theta_{2} \in \mathbb{R}, \theta_1 < \theta_2, \boldsymbol{\eta}_1 \in \mathbb{R}^{w}, \boldsymbol{\eta}_2 \in \mathbb{R}^{w} \right\}$ under $H_1$. 

With a finite sample, one cannot simultaneously control the probability of making a type I error of rejecting $H_0$ when it is true and the probability of making a type II error of accepting $H_0$ when it is false \citep{leh2006}. It is customary to minimize the type II error rate, which is one minus $power$, subject to an upper bound $\alpha$ on the type I error rate \citep{leh2006}. However, in the composite hypothesis testing, the uniformly most powerful (UMP) level $\alpha$ test or the UMP unbiased level $\alpha$ test may not exist when there is no single test that is the most powerful one among all tests or all unbiased tests under every distribution in $\mathcal{F}_1$ under $H_1$. For example, on testing the location parameter with a single observation from Cauchy distribution, the functional form of the most powerful test depends on the underlying location parameter and hence no UMP test exists \citep{leh2006}. Even within the exponential distribution family, it can be challenging to prove the existence and further characterize such optimal test in a finite sample \citep{leh2006}. For example in the famous $\textit{Behrens-Fisher problem}$ of testing the equality of the means from two normal distributions with unknown variances and unknown ratio of variances, the {\it Welch approximate} t-{\it test} is a popular approximate solution in practice \citep{wel1951}. 

In this article, we propose a novel two-stage method to perform hypothesis testing with the aim of increasing power in finite-sample problems when the theoretically optimal test does not exist or is hard to characterize.

\subsection{Motivated problem: simple hypothesis testing}
We motivate our proposed method from the Neyman–Pearson Lemma in a simple hypothesis testing problem on $\theta$ with known nuisance parameters $\boldsymbol{\eta}_0$ using a single group of data $\boldsymbol{x}$ with pmf or pdf $f(\boldsymbol{x} ; \theta, \boldsymbol{\eta}_0)$. The objective is to test the following simple $H_0^\prime$ against simple $H_1^\prime$,
\begin{align}
H_0^\prime: & \:\: \theta = t_0, \:\:\:\:\:\: \text{versus} \:\:\:\:\:\: H_1^\prime: \:\: \theta = t_1, \label{H_0_eg_simple}
\end{align}
where $t_0<t_1$ are two constants. We define a test function $\phi(\boldsymbol{x}) = 1$ if $H_0^\prime$ is rejected, and $\phi(\boldsymbol{x}) = 0$ otherwise. The rejection region is given by $R(\phi) = \left\{\boldsymbol{x}: \phi(\boldsymbol{x}) = 1 \right\}$. Based on the Neyman–Pearson Lemma, a test $\phi(\boldsymbol{x})$ that satisfies
\begin{align}
& \phi(\boldsymbol{x}) = I\left\{f(\boldsymbol{x} ; t_1, \boldsymbol{\eta}_0) > c^{\dagger}f(\boldsymbol{x} ; t_0, \boldsymbol{\eta}_0) \right\} \:\:\:\:\: \text{and} \:\:\:\:\: \alpha = \mathrm{Pr}_{H^\prime_0}\left\{\boldsymbol{x} \in R(\phi) \right\}, \label{equ_simple_np} 
\end{align}
for some $c^{\dagger} \geq 0$ and $I(\cdot)$ as an event indicator function, is a UMP level $\alpha$ test \citep{leh2006}. For example, when data $\boldsymbol{x}$ of size $n$ are assumed to follow a normal distribution with unknown mean $\theta$ and known variance $\sigma^2_0$, the UMP level $\alpha$ test of (\ref{H_0_eg_simple}) is the $z$-test. It rejects $H^\prime_0$ if $\widehat{\theta}(\boldsymbol{x}) >  {\sigma_0} z_{1-\alpha}/\sqrt{n}+t_0$, where $\widehat{\theta}(\boldsymbol{x})$ is the sample mean as a sufficient statistic for $\theta$,  $z_u = \Phi^{-1}(u)$, and $\Phi(\cdot)$ is the cumulative distribution function for the standard normal distribution. 

As an alternative, we formulate the hypothesis testing in the context of a binary classification problem to categorize whether $\boldsymbol{x}$ is sampled from $H^\prime_1$ or from $H^\prime_0$. We introduce a latent variable $y \in \{0,1\}$ indicating where $\boldsymbol{x}$ is drawn. Given $y = k$, the pdf or pmf is equal to $f(\boldsymbol{x} ; t_k, \boldsymbol{\eta}_0)$, for $k = 0,1$. Therefore, the rejection region in (\ref{equ_simple_np}) can be expressed as 
\begin{align}
f(\boldsymbol{x} ; t_1, \boldsymbol{\eta}_0) > c^{\dagger}f(\boldsymbol{x} ; t_0, \boldsymbol{\eta}_0) \iff & \frac{f(\boldsymbol{x} ; t_1, \boldsymbol{\eta}_0)\mathrm{Pr}(y=1)}{f(\boldsymbol{x} ; t_0, \boldsymbol{\eta}_0)\mathrm{Pr}(y=0)} > c^\prime \nonumber \\
\iff & \frac{\mathrm{Pr}(y=1|\boldsymbol{x}, \boldsymbol{\eta}_0)}{1-\mathrm{Pr}(y=1|\boldsymbol{x}, \boldsymbol{\eta}_0)} > c^\prime \nonumber \\
\iff & d\left(\boldsymbol{x}; \boldsymbol{\eta}_0  \right) > \log(c^\prime) \equiv c, \label{equ:dnn_stats}
\end{align} 
where $\Leftrightarrow$" reads if and only if, $d\left(\boldsymbol{x}; \boldsymbol{\eta}_0 \right) = \mathrm{logit} \left\{\mathrm{Pr}(y=1|\boldsymbol{x}, \boldsymbol{\eta}_0)\right\}$, and $\mathrm{logit}(u) =\mathrm{log} \left[ u/(1-u) \right]$ for some constants $c^{\dagger}$, $c^{\prime}$ and $c$. A larger value of $d\left(\boldsymbol{x}; \boldsymbol{\eta}_0 \right)$ indicates that $\boldsymbol{x}$ is more likely to be drawn from $H^\prime_1$ as compared to $H^\prime_0$. The constant $c$ in (\ref{equ:dnn_stats}) is computed to control the type I error rate at $\alpha$,
\begin{equation}
\label{equ:decide_cutoff}
\mathrm{Pr}_{H^\prime_0}\left\{d\left(\boldsymbol{x}; \boldsymbol{\eta}_0 \right) > c \right\} = \alpha.
\end{equation}
Based on the sufficient conditions of the Neyman-Pearson Lemma, any test that satisfies (\ref{equ:dnn_stats}) and (\ref{equ:decide_cutoff}) is most powerful for testing the simple null $H^\prime_0$ at level $\alpha$.

Before considering the composite hypothesis testing in (\ref{H_0_eg}), we first define a set of sufficient statistics $\boldsymbol{t}^{(s)} = \left\{\widehat{\boldsymbol{\theta}}(\boldsymbol{x}_1), \widehat{\boldsymbol{\theta}}(\boldsymbol{x}_2), \widehat{\boldsymbol{\eta}}(\boldsymbol{x}_1), \widehat{\boldsymbol{\eta}}(\boldsymbol{x}_2) \right\}$ of dimension $w^{(s)}$, where $\widehat{\boldsymbol{\theta}}(\boldsymbol{x}_j)$ and $\widehat{\boldsymbol{\eta}}(\boldsymbol{x}_j)$ are sufficient statistics of their true parameters $\theta_j$ and $\boldsymbol{\eta}_j$ based on data $\boldsymbol{x}_j$ from the distribution function $f({x} ; \theta_j, \boldsymbol{\eta}_j)$, for group $j=1, 2$. The superscript ``$(s)$'' indicates that $\boldsymbol{t}^{(s)}$ is utilized in the formulation of statistics in Section \ref{sec:first_DNN}. In some problems, for example the scale-uniform distribution considered in Section \ref{s:sim_eg_1}, $\widehat{\boldsymbol{\theta}}(\boldsymbol{x}_j)$ can be a vector even if $\theta_j$ is a scalar.

A direct generalization from the simple hypothesis testing (\ref{H_0_eg_simple}) to the composite hypothesis testing (\ref{H_0_eg}) is challenging when the likelihood ratio in (\ref{equ:dnn_stats}) depends on unknown parameters $\theta_1$, $\theta_2$, $\boldsymbol{\eta}_1$ and $\boldsymbol{\eta}_2$, because statistics are functions of only data. Following the formulation of $d\left(\boldsymbol{x}; \boldsymbol{\eta}_0  \right)$ in (\ref{equ:dnn_stats}), we intend to identify a statistic $d\left\{ \boldsymbol{t}^{(s)} \right\}$ for composite hypothesis testing such that 
\begin{equation}
\label{equ:composite_stats}
d\left\{ \boldsymbol{t}^{(s)} \right\} = \mathrm{logit} \left[\mathrm{Pr}\left\{ y=1|\boldsymbol{t}^{(s)}\right\}\right],
\end{equation} 
and then obtain the critical value $c$ to control type I error rates at $\alpha$ under $H_0$, 
\begin{equation}
\label{equ:type_1_error_DNN}
\mathrm{Pr}_{H_0}\left[d\left\{ \boldsymbol{t}^{(s)} \right\} > c \right] = \alpha.
\end{equation}
However, the functional form of $d\left\{ \boldsymbol{t}^{(s)} \right\}$ in (\ref{equ:composite_stats}) may not be known explicitly or is even intractable. Another complication is to study the distribution of $d\left\{ \boldsymbol{t}^{(s)} \right\}$ under $H_0$ in a finite sample in order to calculate the critical value of $c$ in (\ref{equ:type_1_error_DNN}). In the following section, we introduce our two-stage method to first characterize the statistic $d\left\{ \boldsymbol{t}^{(s)} \right\}$ and then estimates the corresponding critical value $c$. We particularly leverage Deep Neural Networks (DNN) in the proposed method and discuss its advantages.

\section{DNN-guided hypothesis testing method}
\label{sec:method}

We first provide a short review on DNN in Section \ref{sec:review_DNN}. Then we illustrate our DNN-guided hypothesis testing method by first approximating the test statistics in Section \ref{sec:first_DNN} and then estimating critical values in Section \ref{sec:second_DNN}. The conduct of the proposed method on observed data is demonstrated in Section \ref{sec:DNN_observed_data}. 

\subsection{Review on Deep Neural Networks (DNN)}
\label{sec:review_DNN}

DNN defines a mapping $\boldsymbol{y} = q(\boldsymbol{t}; \boldsymbol{\psi})$ and learns the value of the parameters $\boldsymbol{\psi}$ that result in the best function approximation of output label $\boldsymbol{y}$ based on input data $\boldsymbol{t}$ \citep{goo2016}. The $\textit{deep}$ in DNN stands for successive layers of representations. The last-layer activation function can be chosen as the $\mathrm{sigmoid}$ ($\mathrm{expit}$) function $\mathrm{sigmoid}(u) = 1/\left[1+\mathrm{exp}(-u) \right]$ for binary classification and the $\textit{linear}$ function for a continuous variable approximation, while the inner-layer activation function is usually the $\textit{Rectified Linear Unit}$ ($ReLU$) function defined by $\mathrm{ReLU}(u) = \mathrm{max}(0, u)$ \citep{cho2018}. In obtaining $\widehat{\boldsymbol{{\psi}}}$ to estimate $\boldsymbol{\psi}$ based on a non-convex loss function, we use RMSProp \citep{hin2012} which has been shown to be an effective and practical optimization algorithm for deep neural networks \citep{goo2016}. 

A proper DNN structure and other hyperparameters are usually chosen by cross validation with $80\%$ as the training data and the remaining $20\%$ as the validation data \citep{goo2016}. We start with an architecture with a relatively large capability and a large number of training epochs in the optimization algorithm to reduce the training error. We then apply regulation approaches to increase the generalizability of the model, for example the dropout technique which randomly sets a number of features of the layer as zeros during training, and the mini-batch approach which stochastically selects a small batch of data in computing gradient in the algorithm. We further propose several structures around this suboptimal solution as the candidate pool and select the final skeleton with the smallest model fitting error. Sensitivity analysis in Section \ref{s:sim_eg_1} shows that the performance of our method is robust to different DNN structures (Web Table 3).

\subsection{Approximating the test statistics via DNN}
\label{sec:first_DNN}

In the first stage, we train a DNN to construct the test statistic $d\left\{ \boldsymbol{t}^{(s)} \right\}$ in (\ref{equ:composite_stats}) by using Monte Carlo samples. Note that $\boldsymbol{t}^{(s)}$ does not require minimal sufficient statistics, but only sufficient statistics, which are usually straightforward to obtain based on parametric assumptions. One can also substitute them by order statistics as trivial sufficient statistics, or a vector of key summary statistics, such as mean, median, standard deviation, sample quantiles, etc. In practice, one may evaluate different choices of $\boldsymbol{t}^{(s)}$ to determine the empirically optimal one. For a composite hypothesis testing problem in (\ref{H_0_eg}), we define $\Theta_1 \subseteq \mathbb{R}$ as a neighborhood of the true value of $\theta_1$. The corresponding notations for $\theta_2$, $\boldsymbol{\eta}_1$ and $\boldsymbol{\eta}_2$ are $\Theta_2 \subseteq \mathbb{R}$, $\boldsymbol{H}_1 \subseteq \mathbb{R}^w$ and $\boldsymbol{H}_2\subseteq \mathbb{R}^w$, respectively. We assume that $\Theta_1$, $\Theta_2$, $\boldsymbol{H}_1$ and $\boldsymbol{H}_2$ are all compact. 

To generate the training data, we first simulate $A$ sets of features $\left\{\theta_{1,a}, \theta_{2,a}, \boldsymbol{\eta}_{1, a}, \boldsymbol{\eta}_{2, a} \right\}_{a=1}^A$ from uniform distributions in their corresponding parameter spaces. Then within each set $a$, for $a = 1, \cdots, A$, we simulate $B_0$ samples under $H_0$ with distribution $f({x} ; \theta_{1, a}, \boldsymbol{\eta}_{1, a})$ for group 1 and $f({x} ; \theta_{1, a}, \boldsymbol{\eta}_{2, a})$ for group 2, and simulate $B_1$ samples under $H_1$ with distribution $f({x} ; \theta_{1, a}, \boldsymbol{\eta}_{1, a})$ for group 1 and $f({x} ; \theta_{2, a}, \boldsymbol{\eta}_{2, a})$ for group 2. In each sample $b$, for $b = 1, \cdots, \left\{A \times (B_0 + B_1)\right\}$, we calculate the vector of sufficient statistics
\newline
$\boldsymbol{t}_b^{(s)} = \left\{\widehat{\boldsymbol{\theta}}(\boldsymbol{x}_{1, b}), \widehat{\boldsymbol{\theta}}(\boldsymbol{x}_{2, b}), \widehat{\boldsymbol{\eta}}(\boldsymbol{x}_{1, b}), \widehat{\boldsymbol{\eta}}(\boldsymbol{x}_{2, b}) \right\}$ based on data $\boldsymbol{x}_{1,b}$ and $\boldsymbol{x}_{2,b}$ from two groups to establish the training data $\left\{\boldsymbol{t}_b^{(s)}\right\}_{b=1}^{A \times (B_0 + B_1)}$. The support of $\boldsymbol{t}^{(s)}$ is denoted as $\boldsymbol{T}^{(s)} \subseteq {R}^{w^{(s)}}$. For a sample with index $b$, we further define a classification label $y^{(s)}_b$ taking value either $0$ or $1$, where the event $\left\{y^{(s)}_b = k \right\}$ indicates a sample being drawn from the distribution under $H_k$, for $k = 0, 1$. 

Next, we train the test statistic DNN (TS-DNN) with the $\textit{ReLU}$ as the inner-layer activation function and the $\textit{sigmoid}$ as the last-layer activation function. In the training process, DNN seeks a solution $\widehat{\boldsymbol{{\psi}}}^{(s)}$ which maximizes the log likelihood function, 
\begin{equation}
\label{equ:mle_DNN}
\widehat{\boldsymbol{{\psi}}}^{(s)} =  \argmax_{\boldsymbol{\psi}^{(s)}} \sum_{b=1}^{A \times (B_0 + B_1)} \bigg( \left\{1-y^{(s)}_b\right\}\log \Big[1- p\left\{\boldsymbol{t}_b^{(s)}; {\boldsymbol{\psi}}^{(s)} \right\} \Big] + y_b^{(s)}  \log  p\left\{\boldsymbol{t}_b^{(s)}; {\boldsymbol{\psi}}^{(s)} \right\} \bigg),  
\end{equation}
where $p\left\{\boldsymbol{t}^{(s)}; {\boldsymbol{\psi}}^{(s)} \right\} = \mathrm{sigmoid} \left[q\left\{\boldsymbol{t}^{(s)}; {\boldsymbol{\psi}}^{(s)} \right\}  \right]$, $q\left\{\boldsymbol{t}^{(s)}; {\boldsymbol{\psi}}^{(s)} \right\}$ is the linear predictor of DNN, and ${\boldsymbol{\psi}}^{(s)}$ is a stack of the weight and the bias parameters from all layers in DNN. Essentially, DNN obtains $\widehat{\boldsymbol{{\psi}}}^{(s)}$ by (\ref{equ:mle_DNN}) such that $\mathrm{sigmoid} \left[q\left\{\boldsymbol{t}^{(s)}; {\widehat{\boldsymbol{\psi}}}^{(s)} \right\}  \right]$ approximates the underlying classification probability $\mathrm{Pr}\left\{ y=1|\boldsymbol{t}^{(s)}\right\} = \mathrm{sigmoid} \left[ d\left\{ \boldsymbol{t}^{(s)} \right\} \right]$ in (\ref{equ:composite_stats}). We use $\textit{sigmoid}$ as the last-layer activation function for TS-DNN to follow the formulation based on equation (\ref{equ:dnn_stats}). Before DNN training, we normalize training data with mean zero and unit standard deviation to mitigate the potential gradient issue of $\textit{sigmoid}$. In Web Table 7, we also evaluate DNN with $\textit{softmax}$ as the last-layer activation function. 

The approximation error of utilizing a DNN with linear predictors $q\left\{\boldsymbol{t}^{(s)}; {\boldsymbol{\psi}}^{(s)} \right\}$ to approximate the objective function $d\left\{ \boldsymbol{t}^{(s)} \right\}$ in (\ref{equ:composite_stats}) is defined by the following uniform maximum error \citep{yar2017},
\begin{equation}
\label{equ:appr_error}
\left\vert d - q \right\vert_{\infty} =  \sup_{\boldsymbol{t}^{(s)} \in \boldsymbol{T}^{(s)}} \left\vert d\left\{ \boldsymbol{t}^{(s)} \right\} - q\left\{\boldsymbol{t}^{(s)}; {\boldsymbol{\psi}}^{(s)} \right\} \right\vert. 
\end{equation}
Many theoretical investigations have been done to show that an underlying DNN $q\left\{\boldsymbol{t}^{(s)}; {\boldsymbol{\psi}}^{(s)} \right\}$ can approximate an objective function $ d\left\{ \boldsymbol{t}^{(s)} \right\}$ in a certain function class to have $\left\vert d - q \right\vert_{\infty} =$ upper bounded by a given tolerance, for example Hölder functions \citep{chen2019, she2019}, functions in Sobolev spaces \citep{mha2996, yar2017}, and Lipschitz-continuous functions \citep{bac2017}. 

We provide some discussion on choosing the training features $\left\{\theta_{1,a}, \theta_{2,a}, \boldsymbol{\eta}_{1, a}, \boldsymbol{\eta}_{2, a} \right\}_{a=1}^A$. If one sets $\theta_{1,a} = \theta_{2,a}$ and $\boldsymbol{\eta}_{1, a} = \boldsymbol{\eta}_{2, a}$, then the distribution family $\mathcal{F}_0$ under $H_0$ is exactly the same with $\mathcal{F}_1$ under $H_1$. The resulting solution will be a random classifier with no practical use. On the other hand, if $\theta_{2,a}$ is way larger than $\theta_{1,a}$, then the classification error goes to zero but the trained DNN loses generalizability when $\theta_{2,a}$ is moderately larger than $\theta_{1,a}$. For a set of given $\left\{\theta_{1,a}, \boldsymbol{\eta}_{1, a}, \boldsymbol{\eta}_{2, a}\right\}$, we suggest choosing a $\theta_{2,a}$ such that our DNN-based method reaches a moderate level of power. This can be approximated by some known tests, for example the Student's {\it t}-test. As further demonstrated in Sections \ref{s:sim} and \ref{s:real}, our method has a satisfactory performance in validations when $\theta_2$ is different from this training magnitude. 

\subsection{Approximating the critical values via DNN}
\label{sec:second_DNN}

In the second stage, we train another DNN to estimate the critical value $c$ in (\ref{equ:type_1_error_DNN}) based on simulated samples under $H_0$. 

We construct the training data as $\left\{\boldsymbol{t}^{(c)}_a \right\}_{a=1}^A$ of dimension $(1+2\times w)$ and size $A$, where $\boldsymbol{t}^{(c)}_a = (\theta_{1, a}, \boldsymbol{\eta}_{1,a}, \boldsymbol{\eta}_{2,a})$ are the design features from the previous section. The superscript ``$(c)$'' in $\boldsymbol{t}^{(c)}$ and other notations indicates that they pertain to the estimation of critical values. Within each feature $\boldsymbol{t}^{(c)}_a$, we further simulate $B^\prime$ null data under $H_0$ with group 1 data from the distribution function $f({x} ; \theta_{1, a}, \boldsymbol{\eta}_{1, a})$ and group 2 data from $f({x} ; \theta_{1, a}, \boldsymbol{\eta}_{2, a})$. Their test statistics based on the TS-DNN in the first stage are computed at $\left\{q\left[\boldsymbol{t}_{b^\prime}^{(s)}; \widehat{\boldsymbol{\psi}}^{(s)} \right]\right\}_{b^\prime=1}^{B^\prime}$. The output label $y^{(c)}_a$, for $a = 1, \cdots, A$, is set at the empirical upper $\alpha$ quantile in those test statistics to satisfy (\ref{equ:type_1_error_DNN}). This procedure is performed in a similar fashion as the parametric bootstrap \citep{efr1994} to construct the null distribution of the statistic $q\left[\boldsymbol{t}^{(s)}; \widehat{\boldsymbol{\psi}}^{(s)} \right]$, and then to obtain the corresponding critical values. Under a general composite null hypothesis $H_0$ in (\ref{H_0_eg}), the critical value $y^{(c)}_a$ may depend on the unknown $\theta_{12}$, $\boldsymbol{\eta}_1$ and $\boldsymbol{\eta}_2$, where $\theta_{12}$ is the common value of $\theta_{1}$ and $\theta_2$ under $H_0$. Therefore, we train a critical value DNN (CV-DNN) $q\left\{\boldsymbol{t}^{(c)}; \widehat{\boldsymbol{\psi}}^{(c)} \right\}$ to estimate $y^{(c)}_a$ by $\textit{linear}$ function as the last-layer activation function, and the mean squared error (MSE) as the loss function. As illustrated in Web Appendix E, the proposed procedure saves computational time as compared with the parametric bootstrap method. 

\begin{figure}[h]
	\centering
	\includegraphics[width=10cm]{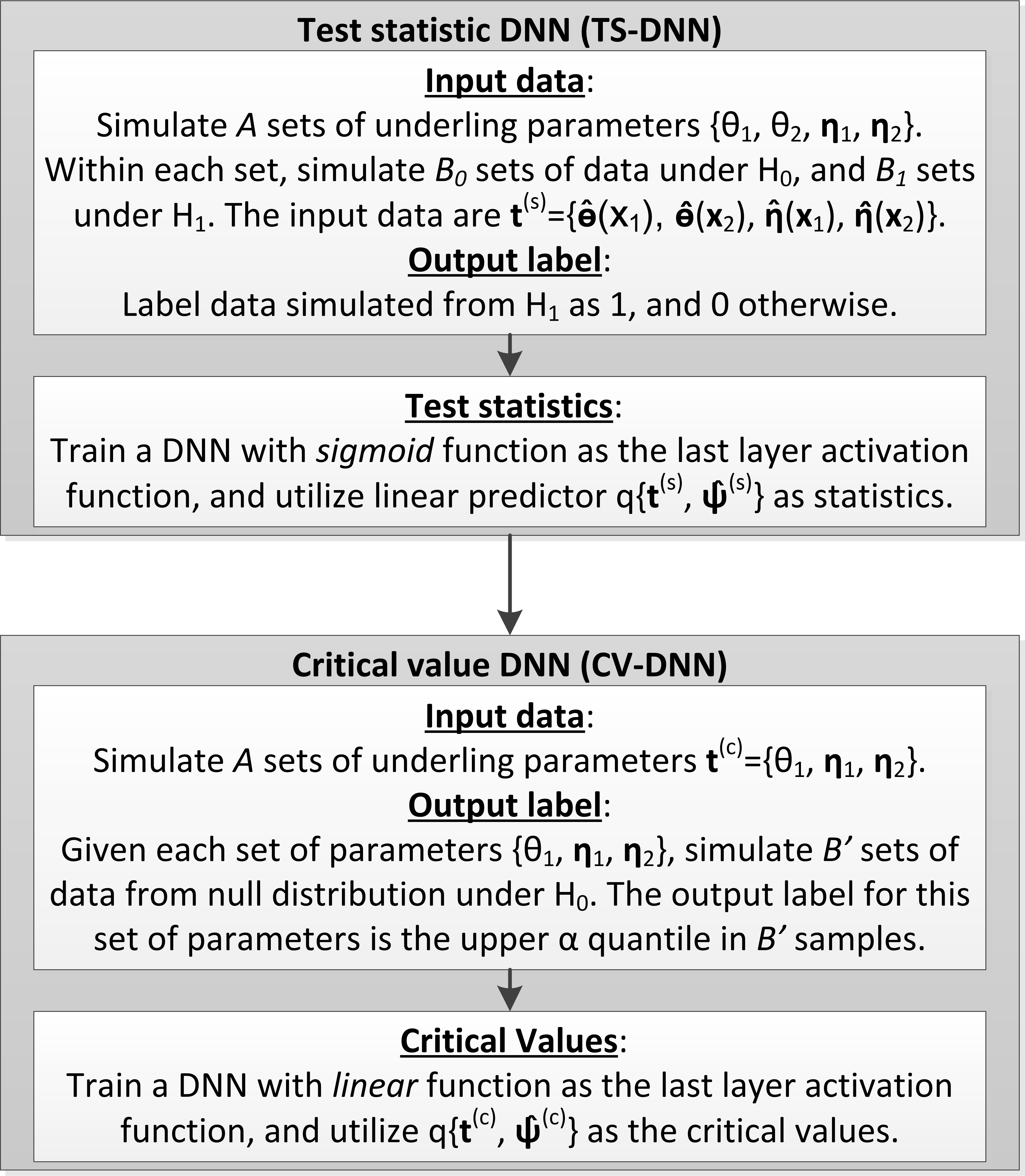}
	\caption{Diagram of the proposed two-stage DNN-based method.}
	\label{F:diagram}
\end{figure}

A diagram is provided to streamline our two-stage method of approximating the test statistics and estimating the critical values training two different DNNs (Figure \ref{F:diagram}). 

At this point, we provide some remarks on utilizing DNN in this framework. First of all, as compared with some simple models like the generalized linear model (GLM) or the linear model (LM), DNN has a stronger functional representation to approximate Lipschitz-continuous functions \citep{bac2017} or functions in Sobolev spaces \citep{yar2017}. As demonstrated in Section \ref{s:sim_eg_1}, if one substitutes the TS-DNN with GLM and substitutes the CV-DNN with LM in our proposed method, then the type I error rate in validation is not controlled at $\alpha$ (Web Appendix D). Secondly, compared to other nonparametric and machine learning methods, DNN provides a scalable way of training on large datasets \citep{goo2016}. The number of simulated training features ($A$, $B_0$ and $B_1$ in Figure \ref{F:diagram}) can be set sufficiently large to give DNN satisfactory performance \citep{goo2016}. We compare its performance with support vector machine (SVM; \citet{bos1992}) and random forest (RF; \citet{ho1995}) with varying hyperparameters and with the same computational cost in Section \ref{s:sim_eg_1} (Web Appendix D). DNN has a more accurate type I error rate control even with reduced numbers of training features. Moreover, DNN is able to automatically identify important and relevant features from data to characterize the objective function without manual feature engineering \citep{cho2018}.

\subsection{Hypothesis testing based on observed data $\widetilde{\boldsymbol{x}}_1$ and $\widetilde{\boldsymbol{x}}_2$}
\label{sec:DNN_observed_data}

Now we are ready to conduct hypothesis testing on (\ref{H_0_eg}) with observed data $\widetilde{\boldsymbol{x}}_1$ from group 1 and $\widetilde{\boldsymbol{x}}_2$ from group 2. We first calculate the input data for the TS-DNN at 
\newline
$\boldsymbol{\widetilde{t}}^{(s)} = \left\{\widehat{\boldsymbol{\theta}}(\widetilde{\boldsymbol{x}}_1), \widehat{\boldsymbol{\theta}}(\widetilde{\boldsymbol{x}}_2), \widehat{\boldsymbol{\eta}}(\widetilde{\boldsymbol{x}}_1), \widehat{\boldsymbol{\eta}}(\widetilde{\boldsymbol{x}}_2) \right\}$ and then compute its test statistic $q\left\{\boldsymbol{\widetilde{t}}^{(s)}; \widehat{\boldsymbol{\psi}}^{(s)} \right\}$. Let $\widetilde{\theta}(\boldsymbol{x})$ and $\widetilde{\boldsymbol{\eta}}(\boldsymbol{x})$ be unbiased estimators for $\theta$ and $\boldsymbol{\eta}$, respectively. The critical value is computed at $q \left\{\widetilde{\boldsymbol{t}}^{(c)}; \widehat{\boldsymbol{\psi}}^{(c)} \right\}$, where $\widetilde{\boldsymbol{t}}^{(c)} = \left\{\widetilde{{\theta}}(\widetilde{\boldsymbol{x}}_{12}), \widetilde{\boldsymbol{\eta}}(\widetilde{\boldsymbol{x}}_1), \widetilde{\boldsymbol{\eta}}(\widetilde{\boldsymbol{x}}_2) \right\}$ and $\widetilde{\boldsymbol{x}}_{12} = (\widetilde{\boldsymbol{x}}_1, \widetilde{\boldsymbol{x}}_2)$. Note that $\widetilde{{\theta}}(\widetilde{\boldsymbol{x}}_{12})$ is an unbiased or consistent estimator of $\theta_{1} = \theta_2$ under $H_0$. Finally, $H_0$ in (\ref{H_0_eg}) is rejected if $ q\left\{\boldsymbol{\widetilde{t}}^{(s)}; \widehat{\boldsymbol{\psi}}^{(s)} \right\} > q \left\{\widetilde{\boldsymbol{t}}^{(c)}; \widehat{\boldsymbol{\psi}}^{(c)} \right\}$, but not rejected otherwise.

\section{Experiments}
\label{s:sim}

\subsection{Scale-uniform distribution}
\label{s:sim_eg_1}

In this section, we consider the scale-uniform distribution $unif\Big([1-k]\theta, [1+k]\theta \Big)$ with the positive parameter of interest $\theta$ and a known design parameter $k \in (0, 1)$, where $unif$ denotes the uniform distribution \citep{gal2016}. This distribution has wide applications, for example the product inventory management in economics \citep{wan2008} and the inverse transform sampling \citep{vog2002}. This distribution is also an example where the distribution family does not satisfy the usual differentiability assumptions leading to the Cram\'er-Rao bound and efficiency of MLEs \citep{leh2006, gal2016}. We demonstrate how to utilize our proposed DNN-based method to assist statistical research.  

With two groups of data $\boldsymbol{x}_1$ and $\boldsymbol{x}_2$ of equal size $n = 20$, we are interested in testing $H_0$ against $H_1$ in (\ref{H_0_eg}) with a one-sided type I error rate $\alpha = 0.05$. Although the Likelihood Ratio Test (LRT) based on the asymptotic Chi-square distribution is not valid \citep{leh2006}, one can construct a LRT type statistic $T_1 = \max(\boldsymbol{x}_2)/\max(\boldsymbol{x}_1)$ because the likelihood ratio is a monotonically increasing function of $\max(\boldsymbol{x}_2)/\max(\boldsymbol{x}_1)$. The critical value of $T_1$ given known $k$ can be computed by simulations, because $T_1$ is a pivotal quantity in the  sense that its distribution is independent from the unknown quantity $\theta$ \citep{leh2006}. Next, we apply our proposed method, denoted as ``DNN'', as a benchmark to evaluate the power performance of $T_1$. 

The neighborhoods of the true values of $\theta_1$ and $\theta_2$ are considered at $\Theta = (0.5, 10) \subseteq \mathbb{R}$, and $K = (0, 1) \subseteq \mathbb{R}$ for $k$. Following the procedures as described in Section \ref{sec:method}, we simulate $A=500$ features with $\theta_{1}$ from $\Theta$ and $k$ from $K$, and further set $\theta_2$ under $H_1$ at a value for our approach to reach approximately $90\%$ power. We choose $B_0 = B_1 = 10^4$. In this case, the training data size for TS-DNN is $A \times (B_0 + B_1) = 10^7$, while the training data size for CV-DNN is $A = 500$. The input vector for the TS-DNN is $\boldsymbol{t}^{(s)} = \left\{ \widehat{\boldsymbol{\theta}}(\boldsymbol{x}_1), \widehat{\boldsymbol{\theta}}(\boldsymbol{x}_2), k\right\}$, where $\widehat{\boldsymbol{\theta}}(\boldsymbol{x}_j) = \left\{ \min(\boldsymbol{x}_j), \max(\boldsymbol{x}_j) \right\}$ are sufficient statistics for $\theta_j$ in group $j=1, 2$ \citep{gal2016}. By cross-validation, the final DNN structure is selected as the one with the smallest validation error from $6$ candidate structures, which are all combinations of the number of layers at $2$ and $3$, and the number of nodes per layer at $50$, $100$ and $150$. The number of epochs is $10$, the batch size is $10^4$, and the dropout rate is set at $0.1$. As illustrated in Section \ref{sec:review_DNN}, the above candidate structure pool is formulated by evaluating a wider and deeper DNN structure with a certain dropout rate and a small batch size introduced to reduce overfitting \citep{cho2018}. Sensitivity analyses in Web Appendix C show that the performance of our method is robust under different properly chosen hyperparameters of DNN, including DNN structure, batch size, dropout rate, last-layer activation function of TS-DNN. Specifically, the power performance is consistent with varying batch size of TS-DNN at $100$ and $1000$. For other problems in general, one can implement cross-validation to choose the empirically optimal batch size, which may be less than $100$. The whole training process is implemented by the R package {\texttt{keras}} \citep{all2020a}. 

In the CV-DNN, the training data is of size $A$ with $\boldsymbol{t}^{(c)} = (\theta_{1}, k)$, and the output is computed on $B^\prime = 10^6$ samples under $H_0$. A number of epochs at $10^3$, a batch size of $10$, and a dropout rate of $0.1$ are utilized in this DNN training. When estimating the critical value, we use the sample mean as an unbiased estimator of $\theta_{1} = \theta_2$ under $H_0$. The number of validation iterations is $10^6$. Unless otherwise specified, the above set-up parameters are utilized throughout this article. Note that the size of training data, for example $A$, can be sufficiently large to give DNN a satisfactory performance, and the number of iterations, for example $B^\prime$, can be increased to improve the precision of numeric calculations. 

In Table \ref{T:sim_1}, we evaluate the performance of our method DNN versus the likelihood ratio based statistic $T_1$, the Student's {\it t}-test, Wilcoxon rank-sum test and the maximum mean discrepancy (MMD) considered in \citet{che2019, kir2020, kub2020, liu2020} on testing means under $4$ scenarios with varying $k$ and varying $\theta_1$. Since MMD is computationally intensive, we utilize $10^4$ simulation iterations in its validation. In each scenario, the first row evaluates the type I error rate under $H_0$, while the other three rows are for power under $H_1$. The value of $\theta_2$ in the third row is the same as that in the training data. The second row captures a lower magnitude and the fourth row considers a higher one. Across all scenarios, all methods have controlled type I error rates at $\alpha = 0.05$. Under $H_1$, DNN is generally more powerful than those alternatives. For example when $k=0.2$, $\theta_1 = 5$ and $\theta_2 = 5.222$, DNN has a power of $80.2\%$ in testing $H_0$, as compared with $67.7\%$ for $T_1$, $31.0\%$ for the Student's {\it t}-test, $29.2\%$ for the Wilcoxon rank-sum test, and $10.8\%$ for MMD. Sensitivity analyses in the Supplemental Materials show that our DNN-based method has a consistent power gain under varying scenarios (Web Appendix A). In Web Appendix B, we also evaluate different choices of $\boldsymbol{t}^{(s)}$ as input data for TS-DNN, for example key summary statistics or order statistics, and observe that $\boldsymbol{t}^{(s)}$ with sufficient statistics has the best power performance in this example. Advantages of DNN over other machine learning methods, e.g., support vector machine (SVM; \citet{bos1992}) and random forest (RF; \citet{ho1995}) are also discussed (Web Appendix D). Additionally, we find that the LRT based statistic $T_1$ has a similar or slightly higher power than DNN when $k=0.8$ and $\theta_1 = 1$, but is less powerful in other scenarios.  

The superior benchmark performance of DNN suggests that maybe a better statistic can be constructed from $T_1$. Since the input data $\boldsymbol{t}^{(s)}$ of TS-DNN contains both $\min(\boldsymbol{x}_j)$ and $\max(\boldsymbol{x}_j)$ as sufficient statistics for $\theta_j$, we can modify $T_1$ with $\min(\boldsymbol{x}_j)$ incorporated to obtain the following statistic $T_2$ as another pivotal quantity given known $k$,
\begin{equation}
\label{equ:sim_1_t2}
T_2 = \frac{w \min(\boldsymbol{x}_2)/(1-k)+(1-w) \max(\boldsymbol{x}_2)/(1+k)}{w \min(\boldsymbol{x}_1)/(1-k)+(1-w) \max(\boldsymbol{x}_1)/(1+k)},
\end{equation}
where $\min(\boldsymbol{x}_j)/(1-k)$ and $\max(\boldsymbol{x}_j)/(1+k)$ are consistent estimators of $\theta_j$ \citep{gal2016}, and $w = \left[ (1-k)^2\right] /  \left[ (1-k)^2 + (1+k)^2\right]$ as an inverse variance weight. To investigate if we can further improve power based on $T_2$, we include it as another input data in $\boldsymbol{t}^{(s)}$ when training TS-DNN, and denote this method as ``DNN-T2''. As can be seen from Table \ref{T:sim_1}, DNN-T2 is more powerful than $T_2$ in most scenarios with a power gain as large as $2.7\%$ when $\theta_1=5$ and $k=0.2$, while $T_2$ is slightly more powerful than DNN-T2 with difference no larger than $0.2\%$ when $\theta_1 = 1$ and $k=0.2$. On the one hand, one may argue that there is essentially limited power improvement by constructing DNN-T2 as compared with $T_2$. This evidence supports that $T_2$ is a feasible statistic with satisfactory performance in practice. On the other hand, our well-trained DNN-T2 itself can also be applied in practice with its available functional form. Moreover, if not satisfying with the power performance of $T_2$ and DNN-T2, one may conduct further research to find a better statistic, for example replacing the consistent estimator of $\theta$ by its unbiased estimator \citep{gal2016}, accommodating correlation between $\min(\boldsymbol{x}_j)$ and $\max(\boldsymbol{x}_j)$ when calculating $w$ in (\ref{equ:sim_1_t2}), etc. The new statistics can also be incorporated to the input data of TS-DNN training to find a better one if possible. 

In this example, our proposed automatic method is utilized as a benchmark to evaluate other statistics which are carefully designed by human intelligence, and is also able to leverage existing statistics to identify a statistic with potentially higher power.

\begin{table}[ht]
	\caption{Type I error rate and power evaluation in the scale-uniform distribution with the method(s) with the highest power highlighted with bold font. }
	\medskip
	\label{T:sim_1}
	\centering
	\begin{threeparttable}[t]
	\begin{tabular}{cccccccccc}
		\toprule
		$k$ & $\theta_1$ & $\theta_2$ &  \multicolumn{7}{c}{Type I error rate ({\it italicized}) / power}  \\
		\cmidrule(r){4-10} 
		&  &  &  DNN-T2\tnote{1} & $T_2$\tnote{2} &  DNN & $T_1$\tnote{3} &  Student's {\it t} & Wilcoxon & MMD \\ 
		\midrule
0.2 &   1 & 1 & \it 4.9\% & \it 5.0\% & \it 5.0\% & \it 5.0\% & \it 5.0\% & \it 4.8\% & \it 4.5\% \\ 
 &    & 1.044 &  79.2\% & \bf 79.4\% &  76.2\% & 67.7\% & 31.0\% & 29.2\% & 11.3\% \\ 
 &    & 1.055 &  91.2\% & \bf 91.3\% &  89.9\% & 83.3\% & 41.5\% & 38.7\% & 15.0\% \\ 
 &    & 1.061 &  94.4\% & \bf 94.5\% &  93.7\% & 88.0\% & 47.1\% & 43.7\% & 17.8\% \\ 
\\
0.2 &   5 & 5 & \it 5.0\% & \it 5.0\% & \it 4.8\% & \it 5.0\% & \it 5.0\% & \it 4.8\% & \it 5.0\% \\ 
 &    & 5.222 & \bf 82.1\% & 79.4\% &  80.2\% & 67.7\% & 31.0\% & 29.2\% & 10.8\% \\ 
 &    & 5.277 & \bf 92.5\% & 91.3\% &  91.3\% & 83.3\% & 41.5\% & 38.6\% & 14.8\% \\ 
 &    & 5.305 & \bf 95.2\% & 94.5\% &  94.5\% & 88.0\% & 47.1\% & 43.7\% & 17.1\% \\ 
\\
0.8 &   1 & 1 & \it 5.0\% & \it 5.0\% & \it 4.7\% & \it 5.0\% & \it 5.0\% & \it 4.8\% & \it 4.6\% \\ 
 &    & 1.178 & \bf 88.1\% & \bf 88.1\% &  87.5\% & 87.7\% & 28.4\% & 26.1\% & 12.4\% \\ 
 &    & 1.222 & \bf 94.9\% & \bf 94.9\% &  94.7\% & 94.7\% & 37.1\% & 33.5\% & 17.5\% \\ 
 &    & 1.244 & \bf 96.7\% & \bf 96.7\% &  96.6\% & 96.5\% & 41.6\% & 37.3\% & 20.9\% \\ 
\\
0.8 &   5 & 5 & \it 5.0\% & \it 5.0\% & \it 4.9\% & \it 5.0\% & \it 5.0\% & \it 4.8\% & \it 4.8\% \\ 
 &    & 5.888 & \bf 88.4\% & 88.1\% &  88.0\% & 87.7\% & 28.3\% & 26.0\% & 12.6\% \\ 
 &    & 6.109 & \bf 95.3\% & 94.9\% &  95.0\% & 94.7\% & 37.0\% & 33.5\% & 17.5\% \\ 
 &    & 6.220 & \bf 96.9\% & 96.7\% &  96.8\% & 96.5\% & 41.6\% & 37.3\% & 20.1\% \\ 
		\bottomrule
	\end{tabular}
	\begin{tablenotes}
\item[1] {DNN-T2 utilizes $T_2$ in (\ref{equ:sim_1_t2}) as another input data of $\boldsymbol{t}^{(s)}$ when training TS-DNN.}
\item[2] {$T_2$ has a functional form in (\ref{equ:sim_1_t2}) with critical value given $k$ determined by simulations. }
\item[3] {$T_1 = \max(\boldsymbol{x}_2)/\max(\boldsymbol{x}_1)$ is a likelihood ratio test (LRT) statistic with critical value given $k$ determined by simulations. }
	\end{tablenotes}
\end{threeparttable}
\end{table}

\subsection{Student's {\it t}-distribution}
\label{s:sim_eg_2}

In this section, we consider a problem of testing the numbers of degrees of freedom in the Student's {\it t}-distribution with two groups of data. In robust estimation and modeling, the {\it t}-distribution provides a useful extension from normality assumption to mitigate the impact of outliers \citep{lan1989, pin2001}. Since the Student's {\it t}-distribution is not an exponential family, we apply our method to find a better testing strategy as compared with common alternatives. 

Two groups of data with equal size $n=200$ are used to test $H_0$ in (\ref{H_0_eg}) with a one-sided type I error rate $0.05$ where $\theta_1$ and $\theta_2$ denote the number of degrees of freedom from two groups, respectively. In this example, we consider a constrained hypothesis testing problem where both $\theta_1$ and $\theta_2$ are within $3$ and $10$. Therefore, we simulate underlying $\theta_1$ and $\theta_2$ from $\Theta = (3, 10) \subseteq \mathbb{R}$ when generating training data for TS-DNN and CV-DNN. Since the sufficient statistics for $\theta$ are not common ones, we consider the training input data for TS-DNN as $\boldsymbol{t}^{(s)} = \left\{ \widehat{\boldsymbol{\theta}}(\boldsymbol{x}_1), \widehat{\boldsymbol{\theta}}(\boldsymbol{x}_2)\right\}$, where $\widehat{\boldsymbol{\theta}}(\boldsymbol{x}_j)$ is a vector of key summary statistics: mean, median, standard deviation, minimum, maximum, first quartile, and third quartile for data $\boldsymbol{x}_j$ from group $j = 1, 2$. The training data for CV-DNN is $\boldsymbol{t}^{(c)} = (\theta_{1})$ as the common $\theta$ under $H_0$, and we set $B^\prime = 10^5$ in this example. In the testing stage, the maximum likelihood estimator of $\theta$ within the constraint $(3, 10)$ is plugged into $\boldsymbol{t}^{(c)}$ to estimate critical values. In Table \ref{T:sim_2}, we compare our DNN method against the one-sided Fligner-Killeen test \citep{fli1976} and the one-sided Levene's test \citep{lev1961} on testing variance, and the one-sided likelihood ratio test (LRT) based on the asymptotic Chi-square distribution with degree of freedom of one \citep{leh2006}. Note that for the Fligner-Killeen test and Levene's test, the one-sided alternative hypothesis is transformed to if the variance of group 1 is larger than that from group 2, because the variance of $t$-distribution $\theta / (\theta-2)$ is a decreasing function of $\theta$. Some tests that are sensitive to normality are not considered due to potential type I error rate inflation, for example the Bartlett's test. We also evaluate another method denoted as DNN-LRT, which incorporate LRT statistic into the training data $\boldsymbol{t}^{(s)}$ of TS-DNN. 

Within each of the two blocks in Table \ref{T:sim_2}, the first row captures the type I error rate, while the next three rows consider the power. The third row evaluates $\theta_2$ under $H_1$ from the training stage. When $\theta_1 = 4$, all five methods have accurate type I error rate controlled at $5\%$, and our DNN-LRT and DNN methods are more powerful than the other three comparators. For example when $\theta_1 = 4$ and $\theta_2 = 7$, DNN has more than $1\%$ power gain than LRT, and over $10\%$ gain than two other tests on variance. When $\theta_1 = 7$, we observe that LRT has a conservative type I error rate at $3.4\%$, and leads to power loss under alternative hypothesis. The reason is that the asymptotic distribution of LRT may not be a single Chi-square distribution when $\theta$ is close to or on the boundary of its parameter space \citep{chen2010}. Without further derivation of the distribution of statistics in finite-sample or even asymptotically, our automatic framework leverages DNN to learn a feasible statistic with satisfactory power and controlled type I error rate. When comparing DNN-LRT and DNN, we observe that DNN-LRT is generally more powerful than DNN with some numerical advantages. This study demonstrates that our framework is general and has the ability to integrate other existing statistics to identify a new test with potentially higher power. 

\begin{table}[ht]
	\caption{DNN-LRT and DNN achieve a higher power than the other three comparators with controlled type I error rates in testing degrees of freedom in $t$-distribution under varying scenarios.}
	\medskip
	\label{T:sim_2}
	\centering
	\begin{threeparttable}[t]
		\begin{tabular}{ccccccc}
			\toprule
			$\theta_1$ & $\theta_2$ &  \multicolumn{5}{c}{Type I error rate ({\it italicized}) / power} \\
			\cmidrule(r){3-7}   &  & \bf DNN-LRT\tnote{1} & DNN & LRT\tnote{2} & Fligner-Killeen & Levene \\ 
			\midrule
4 &   4 & \it 5.1\% & \it 5.1\% & \it 5.0\% & \it 5.0\% & \it 5.0\% \\ 
&   5 & \bf 17.1\% & 16.9\% & 16.3\% & 10.8\% & 13.0\% \\ 
&   6 & \bf 31.5\% & 31.0\% & 30.0\% & 16.8\% & 21.8\% \\ 
&   7 & \bf 44.6\% & 43.9\% & 42.6\% & 22.2\% & 29.9\% \\ 
\\
7 &   7 & \it 4.8\% & \it 4.8\% & \it 3.4\% & \it 4.9\% & \it 4.9\% \\ 
&   8 & \bf 7.8\% & 7.7\% & 5.2\% & 6.7\% & 7.1\% \\ 
&   9 & \bf 11.1\% & 11.0\% & 7.0\% & 8.5\% & 9.3\% \\ 
&  10 & \bf 14.5\% & 14.3\% & 8.7\% & 10.1\% & 11.3\% \\ 
			\bottomrule
		\end{tabular}
		\begin{tablenotes}
\item[1] {DNN-LRT utilizes LRT statistic as another input data of $\boldsymbol{t}^{(s)}$ when training TS-DNN.}
\item[2] {The one-sided likelihood ratio test (LRT) is based on asymptotic Chi-square distribution with one degree of freedom \citep{leh2006}. }
		\end{tablenotes}
	\end{threeparttable}
\end{table}

\subsection{Normal distribution with equal variance assumption}
\label{s:sim_eg_3}

We consider a problem of testing means of two groups of data from the normal distribution $\mathcal{N}(\theta, \sigma^2)$ with equal variance assumption. The Student's {\it t}-test is the known UMP unbiased level $\alpha$ test for testing the composite hypothesis in (\ref{H_0_eg}) \citep{leh2006}. We implement our proposed method in this problem and compare its performance with this theoretically optimal test. 

Since the normal distribution is in a location-scale family, then we set
\newline $\boldsymbol{t}^{(s)} = \left\{ \widehat{{\theta}}(\boldsymbol{x}_2)-\widehat{{\theta}}(\boldsymbol{x}_1), \widehat{\sigma}(\boldsymbol{x}_1), \widehat{\sigma}(\boldsymbol{x}_2)\right\}$ for the TS-DNN and $\boldsymbol{t}^{(c)} = (\sigma)$ for the CV-DNN, where $\widehat{{\theta}}(\boldsymbol{x})$ is the sample mean and $\widehat{{\sigma}}(\boldsymbol{x})$ is the sample standard deviation. In the training stage, we fix $\theta_1$ at $0$, and consider a neighborhood of $\sigma$ at $\Sigma = (0.2, 2) \subseteq \mathbb{R}$. In the validation stage, $\left\{\widehat{\sigma}(\boldsymbol{x}_1)+ \widehat{\sigma}(\boldsymbol{x}_2)\right\}/2$ is plugged into $\boldsymbol{t}^{(c)}$ to compute critical values. 

Table \ref{T:sim_3} shows the type I error rate and power of DNN and the theoretically optimal test Student's {\it t}-test with $n=50$ per group, varying $\theta_1$, $\theta_2$ and $\sigma$. In addition to an accurate type I error rate controlled at $\alpha = 5\%$, DNN achieves a similar power as compared with the Student's {\it t}-test with a deviance not exceeding $0.1\%$ under all scenarios evaluated. Our proposed method well approximates the existing UMP unbiased level $\alpha$ test in this case. 

\begin{table}[ht]
	\caption{DNN reaches the upper power limit from the Student's {\it t}-test as the UMP unbiased level $\alpha$ test. }
	\medskip
	\label{T:sim_3}
	\centering
	\begin{tabular}{ccccc}
		\toprule
		$\theta_1$ & $\theta_2$ & $\sigma$ &  \multicolumn{2}{c}{Type I error rate ({\it italicized}) / power}  \\
		\cmidrule(r){4-5} 
		&  &  &  DNN & Student's {\it t} \\ 
		\midrule
		-0.5 & -0.5 & 1 & \it 5.0\% & \it 5.0\% \\ 
		& -0.1 & &  63.0\% & 63.0\% \\ 
		& 0 & &  79.5\% & 79.5\% \\ 
		& 0.1 & & 90.6\% & 90.6\% \\ 
		\\
		-0.5 & -0.5 & 1.5 & \it 5.0\% & \it 5.0\% \\ 
		& 0.1 & &  62.9\% & 63.0\% \\ 
		& 0.25 & &  79.4\% & 79.5\% \\ 
		& 0.4 & &  90.6\% & 90.7\% \\
		\\ 
		0 & 0 & 1 & \it 5.0\% & \it 5.0\% \\ 
		& 0.4 & &  63.0\% & 63.1\% \\ 
		& 0.5 & &  79.4\% & 79.5\% \\ 
		& 0.6 & &  90.6\% & 90.6\% \\ 
		\\
		0 & 0 & 1.5 & \it 5.0\% & \it 5.0\% \\ 
		& 0.6 & &  62.8\% & 62.9\% \\ 
		& 0.75 & &  79.5\% & 79.6\% \\ 
		& 0.9 & &  90.6\% & 90.6\% \\ 
		\bottomrule
	\end{tabular}
\end{table}

\section{Adaptive clinical trials}

\label{s:real}

\subsection{The Adaptive COVID-19 Treatment Trial (ACTT)}

\label{s:ACTT}

In this section, we apply our method to the Adaptive COVID-19 Treatment Trial (ACTT) to evaluate the efficacy of remdesivir from Gilead Inc. in hospitalized adults diagnosed with COVID-19 \citep{nihcovid}. As illustrated in Section \ref{sec:motivation_two_sample}, adaptive clinical trials are appealing under the pandemic with limited knowledge on COVID-19, because they are capable of accommodating uncertainty during study conduction. As an alternative to existing methods to control type I error rates \citep{bau1994, cui1999}, our DNN-based method builds pre-specified function to seek power enhancement in order to make the adaptive clinical trials more efficient and more ethical. 

In this case study based on ACTT, we consider the sample size reassessment adaptive design for illustrative purposes, which remains the adaptive design most frequently proposed to regulatory agencies for both Food and Drug Administration (FDA) \citep{lin2016} and European Medicines Agency (EMA) \citep{els2014}. For demonstration, we consider a binary endpoint of achieving hospital discharge at Day 14 \citep{nihsimple, gil2020}. The goal is to test $H_0$ versus $H_1$ in (\ref{H_0_eg}) with a one-sided type I error rate $0.05$, where $\theta_1$ is the response rate in the placebo, and $\theta_2$ is from the treatment. The underlying true $\theta_1 = 0.47$ and $\theta_2 = 0.59$ are assumed based on approximations using exponential distributions with median recovery time from the preliminary interim results in \citet{nihinterim}. We consider a two-stage adaptive design with $n^{(1)} = 120$ as the sample size per group in the first stage. A Data and Safety Monitoring Board (DSMB) evaluates unblinded interim data of those $240$ subjects and makes sample size adjustments based on the following rule,
\begin{equation}
\label{adaptive_rule}
n^{(2)} = \left\{
\begin{array}{ll}
n^{(2)}_{min}, \:\:\:\:\:\:\:\:\:\:\:\: \text{if} \:\:\: \widehat{\theta}\left\{\boldsymbol{x}_2^{(1)}\right\} - \widehat{\theta}\left\{\boldsymbol{x}_1^{(1)}\right\} > \theta_{min} \\
n^{(2)}_{max}, \:\:\:\:\:\:\:\:\:\: \text{otherwise}
\end{array}
\right.
\end{equation}
where $\widehat{\theta}\left\{\boldsymbol{x}^{(h)}_j\right\}$ is the sample average, $\boldsymbol{x}^{(h)}_j$ is a vector of observed binary data of size $n^{(h)}$ for group $j$, $j=1, 2$ at stage $h$, $h=1, 2$, and $n^{(2)}_{min}$, $n^{(2)}_{max}$ and $\theta_{min}$ are pre-specified design features. Basically, $n^{(2)}$ in the second stage will be decreased to $n^{(2)}_{min}$ if a promising treatment effect larger than a clinically meaningful difference $\theta_{min}$ is observed, but increased to $n^{(2)}_{max}$ otherwise. Other adaptive measures can also be applied, for example the conditional power \citep{meh2011}. 

We consider a design with $n^{(2)}_{min} = 30$, $n^{(2)}_{max} = 400$, $\theta_{min} = 0.1$, and $\Theta = (0.15, 0.8)$ to cover the underlying $\theta_1 = 0.47$ and $\theta_2 = 0.59$. Our training vector is 
\newline
$\boldsymbol{t}^{(s)} = \left[\widehat{\theta}\left\{\boldsymbol{x}_1^{(1)}\right\}, \widehat{\theta}\left\{\boldsymbol{x}_2^{(1)}\right\}, \widehat{\theta}\left\{\boldsymbol{x}_1^{(2)}\right\}, \widehat{\theta}\left\{\boldsymbol{x}_2^{(2)}\right\}, n^{(2)}\right]$ for the TS-DNN, and $\boldsymbol{t}^{c} = (\theta_{1})$ for the CV-DNN, where $\widehat{\theta}\left\{\boldsymbol{x}\right\}$ is the sample mean of $\boldsymbol{x}$. With observed data in the first stage $\widetilde{\boldsymbol{x}}^{(1)}_{12} = \left\{ \widetilde{\boldsymbol{x}}^{(1)}_{1}, \widetilde{\boldsymbol{x}}^{(1)}_{2} \right\}$, we use $\widehat{\theta}\left\{\widetilde{\boldsymbol{x}}^{(1)}_{12}\right\}$ to estimate the critical value. We evaluate the performance of our DNN-based method versus two existing methods: the inverse normal combination test approach (INCTA; \citet{bau1994, cui1999}) and the empirical test (ET; \citet{ber2010}). The INCTA combines the {\it p}-values from two stages using pre-specified weights, for example equal weights, such that the nominal level can still be applied \citep{bre2009}. The ET approach utilizes the traditional proportional test on the pooled data from two stages and chooses the critical value in the $p$-value scale by a grid search method to control the type I error rate in validation \citep{ber2010}.  

In Table \ref{T:case}(a), we first study the type I error rates under $H_0$ where the common $\theta$ in two groups takes the values $0.37$, $0.47$, $0.57$ and $0.67$ around the underlying $\theta_1 = 0.47$ \citep{nihinterim}. All three methods have type I error rates controlled at $0.05$, where the critical value of ET in the $p$-value scale is $0.032$. In terms of power evaluation, we fix $\theta_1$ in group 1 at $0.47$, and consider varying $\theta_2$ at $0.58$, $0.59$ and $0.6$. Under the true $\theta_2 = 0.59$, DNN consistently has a higher power than two methods, with $6.6\%$ gain as compared with the INCTA and $4.3\%$ gain versus the ET. This superior power performance of DNN is also available under varying designs as shown in Web Appendix F. 

We provide some discussion on the superior power performance of the DNN-based approach. The existing methods INCTA and ET aim at type I error protection, and hence their power may not be optimal. Our proposed method, on the contrary, constructs a test statistic to category whether data come from $H_1$ or from $H_0$ to enhance power in Section \ref{sec:first_DNN}, and further computes its corresponding critical value to control type I error rate in Section \ref{sec:second_DNN}. This formulation also leads to a natural interpretation of our DNN statistic: a measure to optimally category whether the observed data support $H_1$ (the study drug has a better efficacy profile than placebo) or support $H_0$ (there is no treatment effect).

We further calculate the average sample size (ASN) for each method to reach approximately $90\%$ by varying $n^{(2)}_{max}$ in (\ref{adaptive_rule}). DNN requires the smallest ASN at $496$, while $898$ for INCTA and $604$ for ET, demonstrating that our proposed method essentially leads to a more efficient and ethical adaptive clinical trial to evaluate treatment options for COVID-19. To ensure study integrity, regulatory agencies usually require that hypothesis testing strategy to be pre-specified before the current Phase III trial conduct \citep{neu2010}. The two well-trained DNNs from our method can be locked in files to satisfy this requirement. As demonstrated in the R shiny app (link provided in the Section of Supplementary Materials), one can instantly calculate the test statistic from TS-DNN and the critical value from CV-DNN to conduct hypothesis testing based on observed data.  

\begin{table}[!htb]
	\caption{DNN consistently achieves a higher power than INCTA and ET in two adaptive designs with sample size reassessment: the ACTT on COVID-19 and the MUSEC on multiple sclerosis.}
	\label{T:case}
	\begin{threeparttable}[t]
		\begin{subtable}{.5\linewidth}
			\caption{The ACTT}
			\label{T:case_1}
			\centering
			\begin{tabular}{ccccc}
				\toprule
				$\theta_1$ & $\theta_2$ & \multicolumn{3}{c}{Type I error rate} \\ 
				\cmidrule(r){3-5} 
				& & DNN & INCTA$^1$ & ET$^2$  \\
				\midrule
0.37 & 0.37 & 4.9\% & 5.0\% & 4.8\% \\ 
0.47 & 0.47 & 4.9\% & 5.1\% & 4.8\% \\ 
0.57 & 0.57 & 5.0\% & 5.1\% & 4.9\% \\ 
0.67 & 0.67 & 5.0\% & 5.0\% & 4.7\% \\ 
				\\
				\midrule
				$\theta_1$ & $\theta_2$ & \multicolumn{3}{c}{Power} \\ 
				\cmidrule(r){3-5} 
				& & \bf DNN & INCTA$^1$ & ET$^2$  \\
				\midrule
  0.47 & 0.58 & \bf 89.3\% & 84.1\% & 86.2\% \\ 
  0.47 & 0.59 & \bf 93.9\% & 87.3\% & 89.6\% \\ 
  0.47 & 0.60 & \bf 96.5\% & 89.7\% & 92.0\% \\ 
				\bottomrule
			\end{tabular}
		\end{subtable}%
		\begin{subtable}{.5\linewidth}
			\centering
			\caption{The MUSEC}
			\label{T:case_2}
			\begin{tabular}{ccccc}
				\toprule
				$\theta_1$ & $\theta_2$ & \multicolumn{3}{c}{Type I error rate} \\ 
				\cmidrule(r){3-5} 
				& & DNN & INCTA$^1$ & ET$^2$  \\
				\midrule
				0.17 & 0.17 & 4.9\% & 5.0\% & 4.5\% \\ 
				0.27 & 0.27 & 5.1\% & 5.0\% & 4.8\% \\ 
				0.37 & 0.37 & 4.9\% & 5.0\% & 4.8\% \\ 
				0.47 & 0.47 & 4.9\% & 5.0\% & 5.2\% \\ 
				\\
				\midrule
				$\theta_1$ & $\theta_2$ & \multicolumn{3}{c}{Power} \\ 
				\cmidrule(r){3-5} 
				& & \bf DNN & INCTA$^1$ & ET$^2$  \\
				\midrule
  0.27 & 0.39 & \bf 87.4\% & 82.9\% & 83.6\% \\ 
  0.27 & 0.40 & \bf 91.3\% & 85.9\% & 86.4\% \\ 
  0.27 & 0.41 & \bf 93.8\% & 88.2\% & 88.5\% \\ 
				\bottomrule
			\end{tabular}
		\end{subtable} 
	\end{threeparttable}
	\newline
	\newline
	{$^1$ The inverse normal combination test approach \citep{bau1994, cui1999}.
		\newline
		$^2$ The empirical test \citep{ber2010}. }
\end{table}

\subsection{The Multiple Sclerosis and Extract of Cannabis (MUSEC) trial}

\label{s:MUSEC}

We apply our method to the Multiple Sclerosis and Extract of Cannabis (MUSEC) trial, which implemented a sample size adaptive design \citep{zaj2012}. We consider a generic adaptive design with $n^{(1)} = 85$, $n^{(2)}_{min} = 28$, $n^{(2)}_{max}=340$ and $\theta_{min} = 0.1$ in (\ref{adaptive_rule}). The underlying response rates for achieving relief from muscle stiffness after $12$ weeks are assumed as $\theta_1 = 0.27$ for the placebo and $\theta_2 = 0.4$ for the treatment based on results in \citet{zaj2012} with the support $\Theta = (0.05, 0.6)$. 

In Table \ref{T:case}(b), we first evaluate type I error rates under null response rates $0.17$, $0.27$, $0.37$ and $0.47$, and then consider power under the underlying placebo rate $\theta_1 = 0.27$. ET uses a critical value of $0.034$ in the $p$-value scale to preserve type I error rates at $\alpha = 0.05$ within the above range of $\theta$ in validation. Our method has consistently higher power than INCTA and ET under different values of $\theta_2$ in addition to well-controlled type I error rates (Table \ref{T:case}(b)). 

\section{Concluding remarks}
\label{s:discussions}

In this article, we propose a novel two-stage hypothesis testing framework to enhance power in a finite sample. As an important application in the ACTT trial on COVID-19, our method can contribute to a study with a shortened timeline, saved resources, and most importantly, fewer patients involved for ethical consideration. 

Motivated by the ACTT, this article focuses on a two-group comparison with equal sample size. Our method can be readily generalized to a two-group comparison with unequal sample size, paired samples, hypothesis with contrasts, and multiple hypotheses testings. In problems where the potential UMP level $\alpha$ test or the UMP unbiased level $\alpha$ test are hard to characterize, we acknowledge that one can construct a more powerful hypothesis testing strategy by studying the parametric assumption in a given problem; but the next complication is to understand the distribution of the test statistics in a finite sample to compute its critical value with a controlled type I error rate. Our proposed method, on the other hand, provides an automatic learning framework of identifying and characterizing such test statistics and critical values based on DNNs to enhance power. It can be a reference measure to evaluate the performance of other proposed testing strategies based on either analytic derivations or numerical approximations. 

There are some potential limitations of the proposed method. First of all, our approach numerically approximates test statistics and critical values with the aim of power enhancement. The power can be slightly lower than the available theoretical most powerful test, for example in Section \ref{s:sim_eg_3}, or other statistics well-designed by human intelligence for a given problem. Secondly, the interpretation of test statistics constructed by TS-DNN needs further investigation. Based on our current framework, it is interpretated as a measure to maximally category whether observed data come from alternative hypothesis as compared with null hypothesis.


\section*{Acknowledgements}
Tianyu Zhan is an employee of AbbVie Inc. Jian Kang is Professor in the Department of Biostatistics at the University of Michigan, Ann Arbor. Kang’s research was partially supported by NIH grants R01DA048993,  R01MH105561 and R01GM124061 and NSF grant IIS2123777. 

Authors thank the editor Tyler McCormick, an associate editor and two reviewers for their insightful comments which greatly improve this article. 

\section*{Disclosure Statement}

Authors have no conflict of interest to declare.



\section*{Supplementary Materials}


\begin{description}
\item[Supporting information:] Additional simulation results at Section \ref{s:sim} and \ref{s:real} are available in the Supplementary Online Material. 
\item[R code:] The R code is available at GitHub to replicate results in simulation studies and case studies of this article \url{https://github.com/tian-yu-zhan/DNN_Hypothesis_Testing}
\item[R Shiny App:] An R shiny app for the example of ACTT in Section \ref{s:ACTT} is available from \url{https://tianyuzhan.shinyapps.io/dnn_hypothesis_testing/}.
\end{description}

\end{document}